The 14th International Conference on Current and Future Trends of Information and Communication Technologies in Healthcare (ICTH 2024)
October 28-30, 2024, Leuven, Belgium

# Current Trends and Future Directions for Sexual Health Conversational Agents (CAs) for Youth: A Scoping Review

Jinkyung Katie Park[a]*, Vivek Singh[b], Pamela Wisniewski[c]

*[a]Clemson University, 105 Sikes Hall, Clemson, SC 29634, USA*
*[b]Rutgers University, 115 Colllge Avenue, New Brunswick, NJ 08901, USA*
*[c]Vanderbilt University, 2201 West End Avenue, Nashville, TN 37235, USA*

**Abstract**

Conversational Agents (CAs, chatbots) are systems with the ability to interact with users using natural human dialogue. While much of the research on CAs for sexual health has focused on adult populations, the insights from such research may not apply to CAs for youth. The study aimed to comprehensively evaluate the state-of-the-art research on sexual health CAs for youth. Following Preferred Reporting Items for Systematic Reviews and Meta-Analyses (PRISMA) guidelines, we synthesized peer-reviewed studies specific to sexual health CAs designed for youth over the past 14 years. We found that most sexual health CAs were designed to adopt the persona of health professionals to provide general sexual and reproductive health information for youth. Text was the primary communication mode in all sexual health CAs, with half supporting multimedia output. Many sexual health CAs employed rule-based techniques to deliver pre-written expert knowledge on sexual health; yet most sexual health CAs did not have the safety features in place. While youth appreciated accessibility to non-judgmental and confidential conversations about sexual health topics, they perceived current sexual health CAs provided limited sexual health information that is not inclusive of sexual and/or gender minorities. Our review brings to light sexual health CAs needing further development and evaluation and we identify multiple important areas for future work. While the new trend of large language models (LLMs) based CAs can make such technologies more feasible, the privacy and safety of the systems should be prioritized. Finally, best practices for risk mitigation and ethical development of sexual health CAs with and for youth are needed.

---

\* Corresponding author. Tel. +1-864-656-3444 Fax. +1-864-656-3444
  E-mail address: jinkyup@clemson.edu

## 1. Introduction

Conversational Agents (CAs, often called chatbots) are systems with the ability to interact with the users via natural human dialogue [1] (e.g., GPT4, Bing). Driven by advances in the underlying language models, conversational agents have been applied in multiple domains including healthcare [2]. Now, CAs are increasingly used by youth for interactive knowledge discovery on sensitive topics, including sexual health topics [3]. Recent reports show that youth are increasingly experiencing sexual health problems these days. For instance, CDC estimates that youth ages 15-24 account for almost half of the 26 million new sexually transmitted infections (STIs) that occurred in the US in 2018 [4]. Yet, they are hesitant to seek professional help on sexual health due to societal views and attitudes towards the topics themselves and perceived public stigma and embarrassment associated with help-seeking in those topics [5]. With the ability to have human-like interactions with the user, sexual health CAs have been designed and developed to support their unique informational needs related to sexual health topics.

CAs have been studied in the sexual health domain for providing information about general sexual and reproductive health and promoting awareness and adherence to protective medication for sexually transmitted infections (STIs). A review of 57 studies involving sexual health CAs found that users are more likely to disclose sensitive information to chatbots as CAs may be less embarrassing, less stigmatizing, and more private than other sexual health services. At the same time, they showed that the limits of a human-bot relationship can disappoint users and lead to disengagement [6]. A narrative synthesis and meta-analysis of 31 studies on sexual health CAs found that CA-based interventions were effective in adherence to preventative measures. In addition, users showed high levels of user satisfaction and intentions to use but were less satisfied with the quality of CA responses [5]. Overall, existing reviews on sexual health CAs found early evidence that sexual health CA may be effective in improving knowledge and increasing self-efficacy in support of healthy sexual behaviors [5, 6].

Despite the growing body of research on conversational agents (CAs) in the health domain, there remains a significant gap in our understanding of sexual health CAs tailored for adolescents and youth. This demographic represents a unique and critical population segment, not only because they are experiencing significant physical and psychological changes, but also due to their nascent engagement with sexual health information. The nuanced needs of adolescents and youth differ markedly from those of adults, necessitating a specialized approach to sexual health education and support. A systematic review is therefore essential to distill the trends and identify the gaps in current research on sexual health CAs for adolescents and youth. However, to date, no systematic review has been conducted to explore trends in research on sexual health CAs for adolescents and youth. To fill the gap in the literature, we conducted a scoping review of empirical studies that focused on CAs to promote the sexual health of youth. In this review paper, we address the following research questions.

- RQ1: What are the sociotechnical considerations of sexual health CAs for youth?
- RQ2: What are the characteristics and evaluation outcomes of empirical research on sexual health CAs for youth?

In RQ, we explore the interconnection between social and technical factors associated with sexual health CAs for youth in empirical research. Social factors included design considerations such as target audience, health context, and CA goal. Technical factors included computational considerations such as AI technique, communication modality, and knowledge base of sexual health CAs for youth. In RQ, we unpack the characteristics of empirical research (e.g., duration of study, number of participants) and the evaluation outcomes (e.g., dependent measures, strengths and weaknesses of CAs) reported in empirical research on sexual health CAs for youth.

## 2. Methods

Following the Preferred Reporting Items for Systematic Reviews and Meta-Analyses (PRISMA) 2020 statement guidelines [7], a comprehensive systematic review of existing literature on sexual health CAs for youth was conducted. We initially searched the literature with the terms ("conversational agent" OR "chatbot") to explore synonyms of those terms used in the literature. Our initial search informed us of the various alternative terms used to describe conversational agents or chatbots thorough search. Then, we identified four relevant and cross-disciplinary databases in the healthcare domain, including ProQuest, Scopus, Web of Science, and PubMed.

Our final search string consisted of the keywords: ("conversational agent" OR "chatbot" OR "virtual agent" OR "virtual assistant" OR "AI assistant" OR "AI bot" OR "social bot") AND (teen OR adolescent OR youth OR young) AND (sexual). The same search string was used to retrieve articles across four databases. The searches were limited to journal articles, conference papers, and book chapters written in English. The publication date was not specified. The search was conducted in June 2024. The initial search resulted in retrieving 189 articles from the four databases (PubMed ($n = 19$), Web of Science ($n = 17$), Scopus ($n = 32$), and ProQuest ($n = 121$)). After removing the duplicate, we had 156 unique articles.

Next, we screened those 156 articles by title, keywords, and abstract, and this allowed us to remove 137 articles. With the 19 remaining articles, we proceeded with the relevancy coding of the full texts based on the following criteria. We included full research papers that 1) were peer-reviewed. Journal articles and conference proceedings were both included, 2) discussed CAs for providing information/resources/support on sexual health topics, 3) discussed CAs designed for youth, adolescents, or young adults, 4) described CAs that permitted two-way interactions that were fully automated (without human mediation), 5) included empirical results. We excluded papers that 1) were non-full or non-reviewed such as works in progress, extended abstracts, reports, reviews, and meta-analyses, 2) included forms of one-way communication and human-mediated communication, 3) did not consider youth population, 4) did not primarily focus on the CAs (e.g., CAs as one feature of the mobile health apps), 5) did not focus on natural human dialogue as a primary communication mode for two-way interaction (e.g., embodied CAs, facial recognition, etc.), 6) were purely theoretical analysis or a review of existing studies.

Through this relevancy coding process, 11 articles were removed and a set of 8 articles were proceeded for cross-reference. To identify additional relevant papers that were not identified in our initial search, we cross-referenced the citations of 8 articles. We confirmed that there are no additional articles that meet our inclusion criteria, which suggests that we reached a saturation point. The final number of articles that were included in our literature review was eight. Figure 1 presents the data screening process following PRISMA.

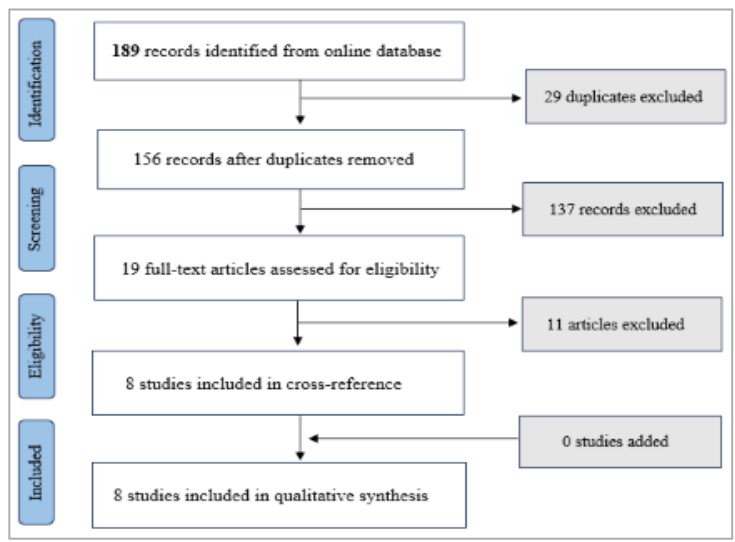

Fig. 1. PRISMA (Preferred Reporting Items for Systematic Reviews and Meta-Analyses) flowchart for the study

We conducted a thematic analysis [8] to identify major themes and trends in our dataset. We leveraged an iterative approach which involved refining the codes as we gained a deeper understanding of the data. For grounded thematic analysis, we analyzed the papers to identify codes for different dimensions aligning with our research questions. We familiarized ourselves with the literature and generated the initial codes. With the initial codes, we coded 20% of the dataset ($n = 2$) and reviewed the codes to ensure that they were representative of our dataset. Once we finalized the codes, the first author carried out the coding, supported by frequent check-ins with the last author for additional expertise on the subject matter. To ensure face validity, the second author quality-checked the coding during the writing of the results. Multiple codes were sometimes assigned to the same paper where necessary. After coding the entire dataset, we grouped our codes into themes: 1) sociotechnical considerations of sexual health CAs, and 2) characteristics and evaluation outcomes of empirical research on sexual health CAs.

## 3. Results

Overall, we included 8 studies in this scoping review [9, 10, 11, 12, 13, 14, 15, 16]. The reviewed studies were published between 2011 and 2023. The majority of the studies were published after 2021, with one publication in 2011; no study was published between 2012 and 2020. The reviewed studies were conducted in diverse places including Bangladesh, Brazil, India, Kenya, Malaysia, Peru, the Netherlands, and the US.

The majority of the reviewed studies included summative evaluation (88%, 7/8), followed by development (75%, 6/8), formative evaluation (63%, 5/8), and design (50%, 4/8). In three studies, research participants were young adults over 18 (38%). In fewer studies, the participants were youth aged between 14-25 or adolescents aged between 12-18 (25%, respectively). In the majority of the studies, participants were general youth (75%, 6/8). In two studies (25%), participants were youth who had higher rates of unintended pregnancy [12] and a higher risk of HIV [14].

*3.1. Sociotechnical characteristics of sexual health conversational agents for youth*

**Most sexual health conversational agents were designed to provide sexual and reproductive health information for general youth:** In the majority of the studies, sexual health CAs were designed for youth and young adults (63%, 5/8). In about half of the reviewed studies, CAs were designed for adolescents. In addition, the majority of the sexual health CAs were designed for general youth populations (63%, 6/8), while only two were designed for at-risk youth (25%). One study designed CA to promote the sexual health of young women with higher rates of unintended pregnancy [12] and the other study developed CA for adolescents who are at a higher risk of HIV [14]. In all studies, the sexual health CAs were designed to provide educational information on general sexual and reproductive health for youth such as STIs, pregnancy, and contraception (100%, 8/8). In one study, sexual health CAs were designed to elicit specific behavioral change, the uptake of pre-exposure prophylaxis (PrEP), one of the HIV prevention methods among adolescent men who have sex with men (AMSM), adolescent transgender women (ATGW) in Brazil [14].

**Sexual health conversational agents were designed to be health professionals with friendly tones:** In 4 studies (50%), the sexual health CA role was mentioned, three of which were health professional-like CAs [10, 11, 12]. For instance, in a study by Wang et al., during the welcoming conversation with users, the CA "SnehAI" introduces itself as Dr. Sneha working in a hospital [11]. In one study, a sexual health CA was designed to be a young and active coach encouraging the adoption of HIV prevention measures [14]. The majority of sexual health CAs were designed to be friendly (63%, 5/8), followed by culturally specific (50%, 4/8). In a couple of studies, sexual health CAs were designed to be gender-specific (25%) and/or simple and factual in their answers (25%). For instance, in a study by Massa et al., the researchers designed the CA "Amanda" with transgender woman persona which provides agile and simple responses using gender-neutral language [14].

**All conversational agents supported text-based input; half of conversational agents supported multimedia output:** In all of the reviewed studies, sexual health CAs supported text as the primary mode of input and output (100%, 8/8). In half of the reviewed studies, CAs supported free textual input. In fewer studies, CAs supported fee text along with quick options (e.g., "yes" or "no") as user input (38%, 3/8) or quick options only (13%, 1/8). In half of the studies (50%, 4/8), sexual health CAs support multi-media outputs. In addition to textual output, the majority of CAs supported image output (50%, 4/8) in the form of an emoji or animated CA avatars. In one study, sexual health CAs also provided outputs in the form of videos and games [11].

**Many conversational agents employed rule-based techniques to deliver pre-written sexual health information:** AI-based computational methods applied in sexual health CAs were addressed in half of the studies, all of which included the rule-based technique (50%, 4/8). With rule-based programming, pre-written sexual health content was delivered based on pre-defined rules. In three studies (38%), the Natual Langauge Processing (NLP) technique was applied along with the rule-based technique [10, 11, 14]. In those studies, quick options are implemented using rule-based methods, but they use techniques such as NLP at certain points in the tree to determine routing to subsequent conversational nodes. There was one study in which CAs relied solely on decisions made by decision trees or rule-based programs [16]. With rule-based CAs, once users choose among the given options (e.g., yes or no, choosing numbered options) the CAs trigger and provide the most relevant pre-written response, hence, there is no space for users to type free text to have conversations.

**Most sexual health content was built upon evidence-based expert knowledge:** In all studies, primary data sources were expert knowledge databases (100%, 8/8). The majority of the sexual health CAs were developed based on expert knowledge of sexual and reproductive health (73%, 6/8). For instance, sexual health content was generated by sexual and reproductive health clinical experts from Population Services International in the US and Kenya [16]. In a few cases (25%, 2/8), researchers collected the most frequently asked question-answer sets about sexual and reproductive health and have them reviewed by clinical experts [10, 12]. For instance, in a study by Bonnevie et al., the authors first worked with Black and Hispanic young women (anonymously) to compile a list of sexual health topics that they were curious about. Then they collaborated with domain experts to come up with answers to those questions to create the knowledge database of sexual health CA [12].

**Most sexual health conversational agents did not have the safety features in place:** The majority of sexual health CAs were without safety features (75%, 6/8). In two papers (25%), safety features of sexual health CAs were addressed. In one study, sexual health CA had a reminder that chatbots are not a replacement for human experts [16]. In another study, sexual health CA provided information on helplines such as the national toll-free helpline endorsed by the Ministry of Health and Family Welfare in India and the national helpline on gender-based violence for the counseling and reporting of incidents [11].

*3.2. Characteristics & evaluation outcomes of empirical research on sexual health conversational agents for youth*

**Most of the research involved user testing to evaluate sexual health conversational agents in the short term:** All empirical studies we reviewed involved user testing to evaluate user engagement and user experience of the sexual health CAs. The number of youth participants in user testing ranged from a minimum of 156 young adults to a maximum of 135,263 online users without demographic information. In three studies (38%), youth interacted with sexual health CAs once for 15-30 minutes [10, 14, 15] for user testing. In five studies (63%), the duration of youth participants' interaction with CAs was not specified. None of the studies indicate that youth interacted with the CAs in the long term. In half of the reviewed studies, researchers interacted with youth populations to gain early feedback on the initial CA design through interviews, focus groups, and design workshops. In those studies, researchers used participatory design approaches such as iterative design [9], community-based participatory research [12], and co-design [16]. In one study, authors conducted an in-lab experiment to evaluate the effectiveness of sexual health CA with 256 participants including adolescents and medical experts in Bangladesh [10].

**Most studies focused on evaluating acceptability and user engagement:** One of the most evaluated outcomes in the empirical research was the *acceptability* of sexual health CAs (50%, 4/8). For instance, a user testing with 234 youth showed that perceived intelligence of sexual health CA "ANA" positively related to youth's intention to adopt the system while anthropomorphism (i.e., humanness) did not [15]. Another most evaluated outcome of sexual health CAs was *user engagement* through analyzing log data of human-CA interaction (50%, 4/8). For instance, to evaluate user engagement with sexual health CA, "SnehAI," the authors analyzed 8,170,879 messages collected over 5 months and demonstrated that almost half of the user messages were personal queries on various sexual health topics such as safe sex, adolescent sexual health, female reproductive health, and mental health [11]. In three studies (38%), the outcome variable was the *effectiveness* of sexual health CAs. For example, by analyzing conversation log data, Rahman et al. found that 85% of the queries by adolescents were answered by "AdolescentBot" demonstrating the potential of sexual health CAs in providing informational support for adolescents [10]. In a couple of studies, the *usability* of sexual health CAs was evaluated (25%). For instance, Massa et al. showed that adolescents perceived Amanda's communication to be straightforward and easy to understand, and liked the language (i.e., abbreviations, emojis), nonbinary expressions, and slang terms used by Amanda [14].

**Sexual health CAs were perceived as accessible and safe places to ask confidential questions:** In five studies (63%, 5/8), the strengths of sexual health CAs for youth were addressed. The most frequently mentioned strengths of sexual health CAs were 24/7 availability/accessibility (50%, 4/8). In some studies, youth liked the anonymity and confidentiality of conversations they could have with sexual health CAs [10, 16] and hence, felt safe and comfortable talking to CAs [14]. In other studies, youth perceived the CAs were easy to navigate [14, 16] and liked the ability to have interactive and engaging conversations with CAs [10, 11]. In addition, studies found that youth liked the responses generated by the CAs as they were friendly and non-judgmental [14, 16] and accurate [10, 16]; some youth liked the gender-neutral avatar character [14] and felt that sexual health CAs helped reduce mental stress related to sexual health concerns [10].

**Sexual health CAs provided limited/repetitive information on general sexual health topics and were not inclusive of information for sexual and/or gender minorities:** In half of the reviewed studies (50%, 4/8), the weaknesses of the sexual health CAs were addressed. The most frequently addressed weaknesses in the reviewed studies were limited/repetitive content provided by CAs [10, 16]. Particularly, youth perceived that sexual health CAs provided limited information on general sexual health topics and were not inclusive of information for sexual and/or gender minorities [10, 16]. In other studies, youth perceived that sexual health CAs lacked an understanding of human input [14] and provided non-human-like responses such as too-fast or rigid responses generated by CAs [9, 14]. Some youth found CA content with jargon or inaccurate/unclear responses generated by CAs hard to understand [14]. In another study, youth expressed difficulties with the free-text input mode as they found it hard to term the sexual health-related queries as well as concerns related to the privacy of their conversation data [10].

**Research ethics such as institutional review and consent were addressed in half of the studies, yet only a few research considered data privacy:** In half of the reviewed studies, the authors explicitly stated that they acquired participant consent and the studies were approved by their institutional review board (50%, 4/8). Beyond institutional review and participant consent which are mandatory in many institutions, in a couple of studies, authors addressed considerations for privacy and/or confidentiality of youth' digital trace data related to sexual health [10, 14]. For instance, in a study by Rahman et al., the authors created dummy accounts on Facebook and randomly assigned one of them to participants to use during user testing to protect their data privacy [10]. In two studies, research ethics were not mentioned [11, 12]. None of the reviewed studies discussed how the authors provided support to promote the safety of youth participants. Also, none mentioned risk mitigation plans and mandated reporting or provided critical reflections on best practices for working with youth in the sexual health context.

## 4. Discussion

**Sexual health CAs that can support diverse youth with a variety of sexual health issues are needed:** In the majority of the reviewed studies, the target audience of sexual health CAs was the general youth. However, in the US, growing numbers of youth are at risk of poor sexual health outcomes including sexual and/or gender minorities, hence, the Centers for Disease Control underscores the importance of addressing the sexual health education needs of LGBTQ+, among other intersectional groups [17]. Therefore, we call for researchers to ensure that sexual health content for minorities in sexual health education is provided by sexual health CAs. Additionally, the majority of the sexual health CAs were designed for older youth populations. More work is needed for younger youth populations (i.e., adolescents), especially those who may be vulnerable to sexual health issues.

**Multimodal communication mode a key to designing sexual health conversational agents that are inclusive of youth with diverse communication needs**: In terms of communication mode, prior research showed that rule-based CAs with quick options are perceived as restricted in offering personalized advice, leading to low trust in the effectiveness of CAs in providing advice on sensitive topics [3]. Therefore, providing options to freely type queries could benefit youth to explore diverse sexual health topics. At the same time, in evaluative research, youth perceived that it was hard to term the queries from scratch. Therefore, it would still be useful to have quick options to choose from or auto-fill features. In addition, an option for voice-based input methods could be beneficial for supporting youth with diverse communication needs. When it comes to output mode, evaluative research confirmed that some

youth found textual information with jargon hard to understand and preferred multi-media content. Therefore, providing sexual health information in multimedia format would help youth engage with sexual health CAs.

**More advanced AI technologies (e.g., LLMs) are needed to provide interactive and engaging sexual health support for youth:** One of the major limitations of sexual health CAs found in evaluative research was limited content and/or responses provided by CAs. This is because many of the sexual health CAs were built upon a rule-based approach in which pre-defined sets of responses are based on domain-specific knowledge. In addition, in the evaluative research we reviewed, youth reported CAs' lack of human language understanding. Recent advancements in large language models (LLMs) are promising in improving the technical immaturity of sexual health CAs with the ability to understand human language in prompts and generate responses beyond pre-written content. Early evidence demonstrated the effectiveness of the LLMs in generating coherent and relevant answers in the health domain [18]. Yet, when applying LLMs in sexual health CAs for youth, the safety of the information provided by those models should be rigorously considered given recent documentation of age-inappropriate and inaccurate content for youth generated by LLMs [19]. Therefore, CA development and implementations should undergo a robust validation process for safety, particularly, with extra care when designing CAs for vulnerable youth.

**Safety should be prioritized when designing sexual health conversational agents for youth:** Overall, most of the empirical research on sexual health CAs was conducted in the short term. Although preliminary evidence shows positive trends in the usability and acceptability of sexual health CAs, long-term evaluative research with more robust research designs is needed to validate their efficacy before their widespread adoption and use. In addition, we noted a lack of established methods for evaluating the safety of sexual health CAs for unintended adverse effects. Evaluation of accuracy and age-appropriateness is critical to ensure the safety of youth interacting with sexual health CAs, yet none of the reviewed studies evaluated such safety aspects of the systems. Therefore, further research is needed to ensure the efficacy and safety of sexual health CAs for youth. Relatedly, we found a concerning trend in the existing literature on sexual health CAs for youth: *most of the studies did not address the ethical aspects of sexual health* CAs. As sexual health topics are sensitive, particularly for youth, further research is needed to establish ethical standards for working with youth, particularly vulnerable youth, to ensure that participating in research does not harm already vulnerable populations.

**Design Guidelines:** Based on our findings and broader implications, we provide the following guidelines for designing youth-centered sexual health CAs.

- **CA role and characteristics**: The CA role could be health professionals or coaches depending on the health context and target audience. Friendly and non-judgemental characteristics of CAs are important to help youth feel comfortable discussing sexual health topics.
- **CA content**: CA content should be age-appropriate, inclusive, and accurate. The content should be based on evidence-based expert knowledge along with inputs from youth.
- **AI technique**: More advanced language models (e.g., LLMs) are needed to provide diverse, context-aware, and personalized sexual health content for youth.
- **Communication mode**: Free-textual inputs with auto-complete/quick options can help youth formulate questions that require domain knowledge. An option for voice-based input can support youth with diverse communication needs. Text along with multimedia output would be beneficial for youth.
- **Safety**: Information on confidentiality and data privacy should be clear and transparent. Safety features such as emergency contacts for imminent risk should be provided upfront and available 24/7.
- **Ethics**: Safety standards and critical reflections on best practices to co-design CAs with the youth population are needed.

## 5. Conclusion

Conversational agents are increasingly used by youth for sensitive topics such as sexual health topics. Trends in research on sexual health CAs designed for youth have been under-explored. In this review paper, we fill an important gap by synthesizing empirical studies on sexual health CAs designed for youth over the last 14 years. With careful consideration of the health context and needs of specific target groups, sexual health CAs can benefit youth. This can be only achieved when engaging with youth from the early design phases to summative evaluation of the systems. In this regard, we call for further investigation of best practices for risk mitigation strategies and ethical development of CAs with and for youth to promote their sexual health.